# DESIGN AND MODELING OF A MEMS-BASED VALVELESS PUMP DRIVEN BY AN ELECTROMAGNETIC FORCE

*Hsien-Tsung Chang, Chia-Yen Lee\*, Chih-Yung Wen*

Department of Mechanical and Automation Engineering,
Da-Yeh University, Chung-hua, Taiwan, 515
E-mail: cy@mail.dyu.edu.tw

## ABSTRACT

A novel valveless micro impedance pump is proposed and analyzed in this study. The pump is constructed of a upper glass plate, two glass tubes, a PDMS (polydimethylsilxane) diaphragm with an electromagnetic actuating mechanism and a glass substrate. The actuating mechanism comprises an electroplated permanent magnet mounted on a flexible PDMS diaphragm and electroplated Cu coils located on a glass substrate. The upper glass plate, PDMS diaphragm and the glass substrate are aligned and assembled to form a micro channel with a compressible section surrounded by a rigid section, creating an acoustic mismatch in the channel. The electromagnetic force between the magnet and the Cu coils causes the diaphragm to deflect and then creates the accumulative effects of wave propagation and reflection at the junction of the compressible and rigid sections. The resulting pressure gradient in the fluid drives the flow from the inlet to the outlet of the micropump. The constituent parts of the electromagnetic actuator, namely the diaphragm, the microcoils, and the magnet are modeled and analyzed in order to optimize the actuator design. The design models are verified both theoretically and numerically and the relationships between the magnetic force, diaphragm displacement, and diaphragm strength are established. The magnitude of the magnetic force acting on the flexible diaphragm are calculated using Ansoft/Maxwell3D FEA software and the resulting diaphragm deflection simulated by ANSYS FEA software are found to agree with the theoretical predictions. Different diaphragm shapes are investigated and their relative strength and flexibility are compared. It is found that a circular PDMS diaphragm represents the most appropriate choice for the actuating mechanism in the micropump. The desired diaphragm deflection of 15 $\mu m$ is obtained using a compression force of 16 $\mu N$, generated by a coil input current of 0.9A. The diaphragm deflection can be regulated by varying the current passed through the microcoil and hence the flow rate can be controlled. The valveless micro impedance pump proposed in this study is easily fabricated and can be readily integrated with existing biomedical chips. The results of the present study provide a valuable contribution to the ongoing development of Lab-on-a-Chip systems.

## 1. INTRODUCTION

The valveless pumping effect now known as the Liebau phenomenon was first reported by Gerhart Liebau in 1954 and was subsequently examined by Borzi [1]. For the Liebau phenomenon, the application of a periodic force at a position which lies asymmetric with respect to the system configuration generates a valveless pumping effect. Rinderknecht *et al.* [2] proposed a new valveless, substrate-free impedance-based micropump driven by electromagnetic actuation. An impedance pump is composed of an elastic section connected at the ends to rigid sections. In the valveless micro impedance pump reported by Wen [3], a PZT actuator was used to deflect a thin film of Ni. The resulting resonance of the diaphragm vibration generated a larger diaphragm deflection, which consequently produced a greater driving pressure. The micropump design developed in this study considers particularly the requirement to achieve the desired flow rate while maintaining a safe operation. The flow rate in the micropump is directly related to the volume change of the channel caused by the deflection of the diaphragm. Therefore, the first stage in the design procedure is to specify the diaphragm deflection required to generate the desired flow rate. The second stage involves designing a micropump structure capable of producing a sufficient actuating force to achieve this desired deflection in a safe manner.

## 2. DESIGN

The valveless micro impedance pump illustrated schematically in Fig. 1 incorporates PDMS walls, a PDMS diaphragm attached to a wet-etched glass upper plate, a permanent magnet electroplated on the upper surface of the diaphragm, and Cu coils electroplated on a lower glass substrate. When a current is passed through the coils, the resulting magnetic force between the coils and the permanent magnet compresses the diaphragm causing a volume change of the channel. This creates an acoustic impedance mismatch in the fluid, which in turn establishes a pressure gradient within the fluid as a result of wave interference. The pressure potential then drives the fluid from the inlet to the outlet. In the present study, the desired membrane deflection is specified as 15 $\mu m$ and the microchannel is assigned similar dimensions to that in Wen's study [3] to construct the models and improve on the designed micropump.

## 2.1. Electromagnetic force

The electromagnetic force between a permanent magnet and a microcoil which are located along the center line of the coil is given by Feustel *et al.* [4]:

$$F_z = M_z \cdot V_m \cdot (\partial B_z / \partial z) \qquad (1)$$

where $M_z$ is the magnetization of the magnet, $V_m$ is the volume of the magnet, and $B_z$ is the flux density produced by the coil in the vertical direction. Equation (1) indicates that the magnetic force is proportional to the change in flux density produced by the coil in the vertical direction. To optimize the actuator performance, the magnet should be placed in such a position that the gradient of the vertical magnetic field is maximized. The magnetic field, $B_z$, and the gradient of the magnetic field, $\partial B_z / \partial z$, are calculated and analyzed in coil design using Ansoft/Maxwell 3D FEA software.

## 2.2. Actuator Component Design
### 2.2.1. Diaphragm design
This study chooses PDMS as the diaphragm material. PDMS has the advantages of good flexibility, excellent biological compatibility and a high yield strength (elastic modulus E = 750 KPa, Poisson's ratio $v = 0.5$, and yield strength $\sigma_y = 130$ KPa [5]). Therefore, the PDMS diaphragm provides a safe and efficient pumping effect even under resonance conditions. In the study, the PDMS diaphragm is in the form of an edge-clamped thin circular plate with a uniform load $q$ exerted over a central circular area. The plate radius is deliberately specified as 1955 $\mu m$ such that its surface area is equivalent to that of the rectangular plate in Wen's study [3]. The maximum deflection occurs at the center of the plate and the equation of displacement was derived by Timoshenko and Woinowsky-Krieger [6], which is:

$$w_{max} = \left(Fc^2 / 16\pi D\right)\left(\kappa^2 - \ln \kappa - (3/4)\right) \qquad (2)$$

where $F = \pi c^2 q$ is the total load acting on the plate, $\kappa = a/c$, where $a$ is the radius of the plate and c is the radius of the loaded area which also is the area "occupied" by the magnet. The limiting force for a circular plate with clamped edges and a uniformly distributed load over its central circular area was derived in the study [6,7], and which is:

$$F_{lmt} = h^2 \sigma_y \left(4\pi k^2 / (6k^2 - 3)\right) \qquad \text{when} \qquad k < 4.5 \qquad (3a)$$

$$F_{lmt} = h^2 \sigma_y \left(8\pi k^2 / (1+v)(12k^2 \log k + 3)\right)$$
when $k \geq 4.5$ \qquad (3b)

where $\sigma_y$ is the yield strength of the diaphragm material. As stated previously, this study specified a desired deflection value of 15 $\mu m$, i.e. the maximum deflection value of the micro impedance pump presented by Wen *et al.*[3]. The force required to achieve this deflection is determined from Equation (2) to be 16 $\mu N$ with $k = 1.6$. In this study, the thickness of the PDMS diaphragm is specified as 80 $\mu m$ since the limiting force of the diaphragm with this thickness is far more than the force of 16 $\mu N$ which is required to achieve the desired deflection of 15 $\mu m$.

### 2.2.2. Magnet design
The magnet considered in this study is a CoNiMnP electroplated permanent magnet with a magnetic coercivity of 47.7 kA/m and a magnetic remanence of 0.2~0.3 T [8]. In determining an appropriate magnet radius, $c$, it is important that the diaphragm is of a sufficient area to move freely and to be operated safely. Therefore, the radius of the current magnet is specified as 1222 $\mu m$ and its thickness as 20 $\mu m$ where having an appropriate value of $\kappa = 1.6$ and the maximum retentivity of 0.3 T. Note that these parameters are deliberately assigned conservative values to ensure that the diaphragm will not collapse during operation.

### 2.2.3. Coil design
The coil design parameters include the inner radius of the coil, the spacing of the turns, and the cross-sectional area of the conductor material. Initially, this study examined the interrelationships between the coil parameters by considering single planar coils with a constant inner radius of 400 $\mu m$, but varying the number of turns, the width of conductor and the spacing of turns. The results yield some interesting findings, e.g. (a) for coils of the same spacing, width and pitch, no appreciable enhancement of the magnetic force is obtained by increasing the number of turns in the coil; (b) for coils with the same spacing, pitch, and number of turns, the magnetic force increases as the coil width is decreased; and (c) for coils with the same width, pitch, and number of turns, the magnetic force reduces as the spacing is increased.

## 3. RESULTS AND DISCUSSION

To achieve the required actuating force of 16 $\mu N$ and operation safely, this study specified a spiral planar coil with the parameters presented in Table 1, also the required diaphragm and magnet with the parameters are present in table 2 and table 3.he paper title (on the first page) should begin 44 mm (1.73 inch) from the top edge of the page, centered, completely capitalized, and in Times 12-point, boldface type. The authors' name(s) and affiliation(s) appear below the title in capital and lower case letters. Papers with multiple authors / affiliations may require more lines.

The vertical magnetic field produced by the coil from the coil plane and its derivative were calculated for coil input currents of 0.2-1.0 A using Ansoft/Maxwell 3D FEA software. In Fig. 2, it can be seen that the maximum gradient of the magnetic field occurs at a point located 620 $\mu m$ above the planar coil. Therefore, the magnet should be located at this position to optimize the electromagnetic actuation effect. Clearly, the value of the magnetic force generated by the actuator depends on the value of the coil input current. Fig. 3 illustrates the variation of the magnetic force calculated by Ansoft/Maxwell3D FEA software for input currents in the range 0.2 to 1.0 A. It is observed that at these values of input current, the magnetic force varies from 4.58 to 18.4 $\mu N$. With the magnet located at the optimal position of 620 $\mu m$ above the planar coil, an input current of 0.9 A is sufficient to produce the actuating force of 16 $\mu N$ required to generate the desired diaphragm displacement of 15 $\mu m$.

ANSYS FEA software was used to model the deflection at the center of the circular diaphragm for magnetic forces in the range 4.58 to 18.4 $\mu N$. It was found that the deflection ranged from 4.2 to 17 $\mu m$. To validate the models established for the actuator components, the theoretical results for the deflection at the center of the diaphragm were calculated using Equation (2). Fig. 4 plots the theoretical and simulation results for the variation of the diaphragm deflection with the magnetic force. It is apparent that a good agreement exists between the two sets of results. When choosing a suitable diaphragm for the current actuator, this study considered a circular plate, a square plate and a rectangular plate. Note that the various plates were of equivalent dimensions (same surface areas and same thickness) and had similar material properties. Fig. 5 plots the variation in the diaphragm displacement with the applied load for each type of diaphragm. The ANSYS FEA simulation results for the plate deflection under an applied force of 18.4 $\mu N$ are presented in Fig. 6. It can be seen that under this magnitude of applied force, the circular plate undergoes a significantly greater deflection than the other plates. In other words, the circular plate is more readily deflected than the square or rectangular plates.

## 4. CONCLUSIONS

(1) This study has designed and modeled a novel valveless micro impedance pump incorporating an asymmetric force chamber, a flexible polymer membrane, and a permanent magnetic actuator. The parameters of the actuator components have been optimized.

(2) A novel design method has been proposed for the micro impedance pump. This method enhances the micropump performance while maintaining a safe operation. The corresponding analysis models have been verified theoretically and numerically.

(3) The permanent magnetic actuator presented in this study is capable of producing a sufficiently large displacement of the biocompatible PDMS membrane.

## ACKNOWLEDGEMENTS

The authors would like to thank the financial supports provided by the National Science Council in Taiwan (NSC-94-2211-E-212-009 and NSC 94-2218-E-006-045) and the National Center for High-Performance Computing (NCHC) for access to the Ansoft/Maxwell 3D software.

Table 1. Parameters of designed spiral planar coil.

| Parameter | Material | Width | Pitch | Spacing |
|---|---|---|---|---|
| Data | Cu | 25μm | 20μm | 20μm |
| Parameter | Turns | Inner radius | Outer radius | Resistance ( at 20℃ ) |
| Data | 10 | 1250μm | 1725μm | 3.23Ω |

Table 2. Parameters of designed diaphragm.

| Parameter | Material | Radius | Thickness | Limiting force |
|-----------|----------|--------|-----------|----------------|
| Data | PDMS | 1955μm | 80μm | 377μN |

Table 3. Parameters of designed magnet

| Parameter | Material | Radius | Thickness | Remanence(Br) |
|-----------|----------|--------|-----------|---------------|
| Data | CoNiMnP | 1222μm | 20μm | 0.3T |

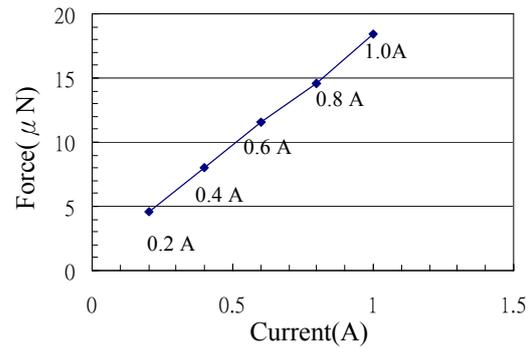

Figure 3 Variation of magnetic force generated by designed actuator with input currents in range 0.2 to 1.0A.

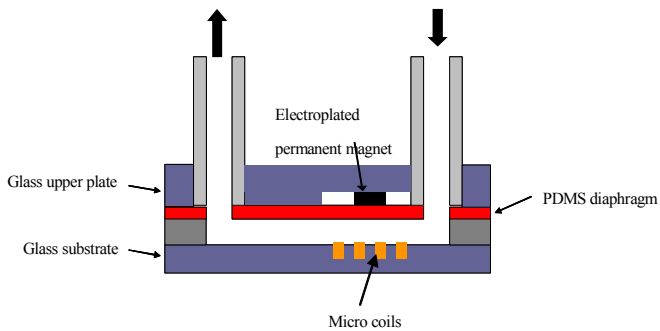

Figure 1 Schematic illustration of valveless micro impedance pump with PDMS diaphragm driven by electromagnetic force.

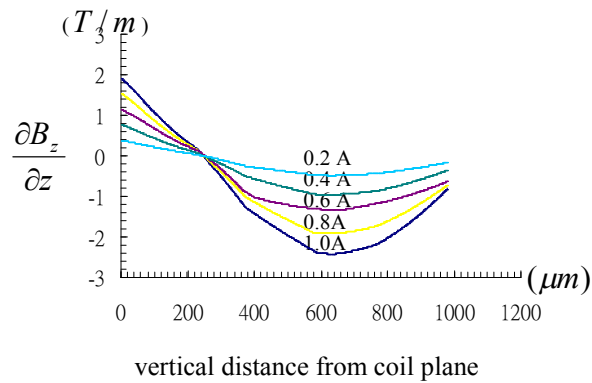

vertical distance from coil plane

Figure 2 Variation of $\partial B_z / \partial z$ with vertical distance from coil plane for designed coil with input currents in range 0.2 to 1.0 A.

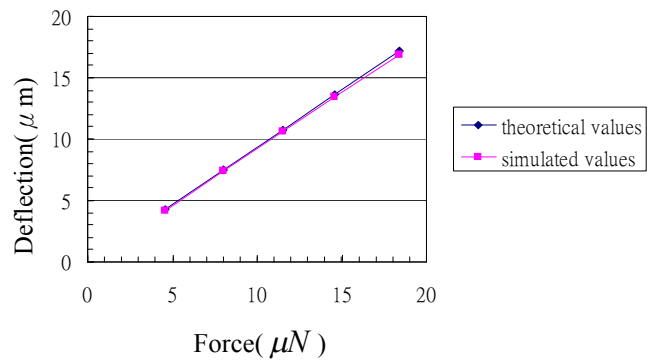

Figure 4 Theoretical and simulation results obtained for deflection at center of diaphragm with magnetic forces given in Fig.3.

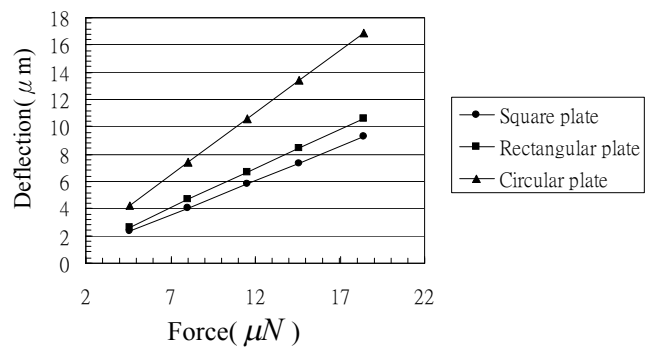

Figure 5 Variation of deflection at center of different plates with applied magnetic force.

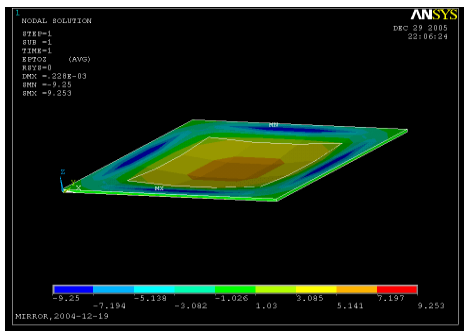

(a) square plate

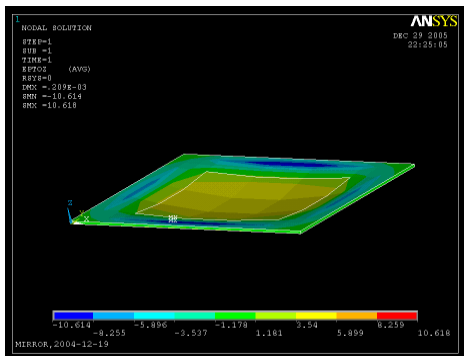

(b) rectangular plate

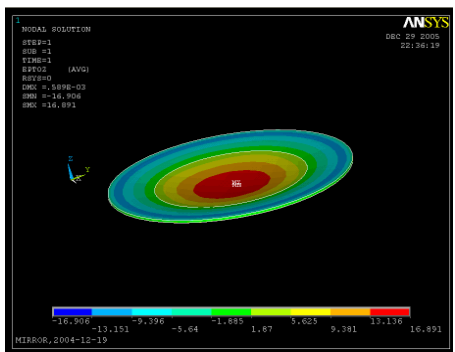

(c) circular plate

Figure 6 Simulation results for deflection of different plates under actuating force of 18.4 $\mu N$ .

## REFERENCES

1. A. Borzi and G. Propst, "Numerical investigation of the Liebau phenomenon," *Math. Phys.* 54, pp. 1050-1072, 2003.

2. D. Rinderknecht, A.I. Hickerson and M. Gharib, "A valveless micro impedance pump driven by electromagnetic actuation," *J. Micromech. Microeng.* 15, pp. 861-866, 2005.

3. C.Y. Wen, C.H. Cheng, C.N. Jian, T.A. Nguyen, C.Y. Hsu and Y.R. Su, "A valveless micro impedence pump driven by PZT actuation," *Materials Science Forum* 505-507, pp. 127-132, 2006.

4. A. Feustel, O. Krusemark and J. Muller, „Numerical simulation and optimization of planar electromagnetic actuators" *Sensors and Actuators A* 70, pp. 276-282, 1998.

5. K.M. Choi and J. A. Rogers, "A Photocurable Poly(dimethylsiloxane) Chemistry Designed for Soft Lithographic Molding and Printing in the Nanometer Regime," *J. AM. CHEM. SOC.* 125, pp. 4060-4061, 2003.

6. Timoshenko, S. and Woinowsky-Krieger, S., *Theory of Plates and shells* 2nd Ed, McGraw-Hill, New York, 1977.

7. Chen, W.F. and Han, D.J., *Plasticity for Structural Engineers* Gau Lih Book Co., LTD, Taiwan, 1995.

8. T.M.Liakopoulos, W.J. Zhang, G.H. Ahn, and Liakopoulos, M. Trifon, "Electroplated Thick CoNiMnP Permanent Magnet Arrays For Micromachined Magnetic Device Applications," IEEE MEMS Proc., pp. 79-84, 1996.